\documentclass{article}
\usepackage[preprint]{spconfa4}
\usepackage{amsmath,amssymb,graphicx}
\usepackage[colorlinks=false,pdfborder={0 0 0}]{hyperref} 
\usepackage{units}
\usepackage[all=normal,mathspacing]{savetrees} 
\usepackage{balance}

\usepackage{url, hyperref} 

\usepackage{tikz,pgfplots}
\usetikzlibrary{decorations.pathreplacing,plotmarks}
\pgfplotsset{compat=1.7}
\usepackage{textcomp}
\usetikzlibrary{shapes,arrows}
\usetikzlibrary{arrows.meta}
\usepackage{mathtools}
\usepackage{environ}
\makeatletter
\tikzset{%
  block/.style    = {draw, thick, rectangle, minimum height = 3em,
    minimum width = 3em},
  sum/.style      = {draw, circle, node distance = 2cm}, 
    input/.style    = {coordinate}, 
  output/.style   = {coordinate} 
}
\newsavebox{\measure@tikzpicture}
\NewEnviron{scaletikzpicturetowidth}[1]{%
  \def\tikz@width{#1}%
  \begin{lrbox}{\measure@tikzpicture}%
  \BODY
  \end{lrbox}%
  \pgfmathparse{#1/\wd\measure@tikzpicture}%
  \BODY
}
\usetikzlibrary{calc}
\def\centerarc[#1](#2)(#3:#4:#5)
    { \draw[#1] ($(#2)+({#5*cos(#3)},{#5*sin(#3)})$) arc (#3:#4:#5); }

\usepackage[noadjust]{cite}

\newcommand{\1}{\mathbf{1}}
\newcommand{\e}{\mathbf{e}}
\newcommand{\0}{\mathbf{0}}
\newcommand{\I}{\mathbf{I}}
\DeclareMathOperator*{\argmax}{argmax}
\DeclareMathOperator*{\argmin}{argmin}

\newcommand{\s}{\mathbf{s}}

\newcommand{\D}{\mathbf{D}}
\newcommand{\G}{\mathbf{G}}

\newcommand{\U}{\mathbf{U}}

\newcommand{\Mbf}{\mathbf{M}}

\newcommand{\pbf}{\mathbf{p}}
\newcommand{\Pbf}{\mathbf{P}}
\newcommand{\Real}{\mathbb{R}}
\newcommand{\dbf}{\mathbf{d}}

\newcommand{\m}{\mathbf{m}}

\newcommand{\ji}{\jmath}

\DeclareMathOperator{\EX}{\mathbb{E}}

\copyrightnotice{978-1-6654-6867-1/22/\$31.00~\copyright2022 IEEE}
                   
\title{3D Single Source Localization Based on Euclidean Distance Matrices}
\name{Klaus Br{\"{u}}mann, Simon Doclo \thanks{This work was funded by the Deutsche Forschungsgemeinschaft (DFG, German Research Foundation) under Germany's Excellence Strategy - EXC 2177/1 - Project ID 390895286 and Project ID 352015383 - SFB 1330 B2.
}}
\address{Department of Medical Physics and Acoustics and Cluster of Excellence Hearing4all, \\
University of Oldenburg, Germany}
\begin{document}
\ninept
\maketitle
\begin{abstract}
A popular approach for 3D source localization using multiple microphones is the steered-response power method, where the source position is directly estimated by maximizing a function of three continuous position variables. 
Instead of directly estimating the source position, in this paper we propose an indirect, distance-based method for 3D source localization. 
Based on properties of Euclidean distance matrices (EDMs), we reformulate the 3D source localization problem as the minimization of a cost function of a single variable, namely the distance between the source and the reference microphone. 
Using the known microphone geometry and estimated time-differences of arrival (TDOAs) between the microphones, we show how the 3D source position can be computed based on this variable. 
In addition, instead of using a single TDOA estimate per microphone pair, we propose an extension that enables to select the most appropriate estimate from a set of candidate TDOA estimates, which is especially relevant in reverberant environments with strong early reflections. 
Experimental results for different source and microphone constellations show that the proposed EDM-based method consistently outperforms the steered-response power method, especially when the source is close to the microphones. 
\end{abstract}
\section{Introduction} 
\label{sec:intro}
The location of a speech source, relative to some microphones (e.g., in mobile phones or hearing aids), is a widely used spatial feature for speech enhancement or speaker extraction. 
Often, the localization constitutes estimating the source direction of arrival using compact microphone arrays, where it can be assumed that the source is in the far field. 
In this paper, we focus on 3D localization, where the far field assumption is not made, i.e., using spatially distributed microphones of an acoustic sensor network. 

Source localization methods \cite{dibiase2000high, madhu2008acoustic, huang2008time, pertila2018multichannel} can be broadly categorized into direct (one-step) and indirect (two-step) approaches. 
The steered-response power with phase transform (SRP-PHAT) method \cite{dibiase2000high} is a direct approach, which exploits the generalized cross-correlations \cite{knapp1976generalized} between all microphone pairs and can be interpreted as a delay-and-sum beamformer, steered towards all possible 3D source positions, and has gained much popularity due to its robustness against noise and reverberation. 
A drawback is that it requires the optimization of three continuous position variables, for which in practice a discrete 3D grid search is used. 
Various methods have been proposed to reduce the computational complexity while achieving comparable localization performance \cite{do2007fast, cobos2010modified, nunes2014steered,garcia2021analytical, dietzen2021low}. 

Instead of directly estimating the source position, in this paper we propose an indirect estimation method based on a Euclidean distance matrix (EDM) \cite{torgerson1952multidimensional, dokmanic2015euclidean}, containing both the distances between the microphones (assumed to be known) and the (unknown) distances between the source and the microphones. 
We propose to decompose the unknown distances into the distance between the source and the reference microphone and a distance component which is proportional to the time-differences of arrival (TDOAs) between the reference microphone and the other microphones. 
Assuming estimates of the TDOAs to be available allows us to formulate the EDM and the related Gram matrix as a function of a single variable, representing the distance between the source and the reference microphone. 
Exploiting the rank property of the Gram matrix, we propose to minimize a cost function which depends on this variable. 
The estimted relative source position can be reconstructed from the Gram matrix, which minimizes the cost function, and can then be aligned to the estimated source position using orthogonal Procrustes analysis \cite{schoenemann1964solution,dokmanic2015euclidean}. 
Since in reverberant environments early reflections may result in large TDOA estimation errors, we propose a method to select the best TDOA estimate from a set of multiple candidate TDOA estimates, based on the same rank property of Gram matrices. 

Experimental results for different source and microphone constellations in noisy and reverberant environments show that the proposed EDM-based 3D source localization method outperforms SRP-PHAT and results in significantly smaller estimation errors when the source is close to the microphones. 
Furthermore, we show that the proposed TDOA selection method leads to a reduction in the number of large localization errors. 

\section{Source Localization Using GCC-PHAT} 
\label{sec: SRP-PHAT}

We consider a reverberant and noisy acoustic environment with a single static speech source and a spatially distributed microphone array with $M>3$ microphones, where $\m_m \in \Real^3$ denotes the position of the $m$-th microphone. 
The aim is to estimate the source position $\s \in \Real^3$ relative to the microphone positions $\Mbf = [\m_{1}, \dots, \m_{M}]$, which are assumed to be known. 
Assuming synchronized microphones and free field transmission, i.e., no object or head between the source and the microphones, the TDOA of the direct speech component between the $i$-th and $j$-th microphone is equal to $\tau_{i,j}(\s) = (||\s-\m_{i}|| - ||\s-\m_{j}||)f_s/\nu$, with $f_s$ the sampling frequency and $\nu$ the speed of sound. 

A common approach to estimate the TDOAs between the microphone pairs is based on the time-domain generalized cross correlation with phase transform (GCC-PHAT) function \cite{knapp1976generalized, velasco2016proposal, zhang2008does}, defined between microphone $i$ and $j$ as 
\vspace*{-1.5 mm}
\begin{equation}
    \xi_{i,j}(\tau) \; = \int_{-\omega_{0}}^{\omega_{0}} \psi_{i,j}^{}(\omega) e^{\ji \omega \tau} d\omega \; ,
    \label{eq: IFT of GCC-PHAT}
\vspace*{-1.2 mm}
\end{equation}
with radian frequency $-\omega_{0} \leq \omega \leq \omega_{0}$ and time lag $\tau$. 
The frequency-domain GCC-PHAT function $\psi_{i,j}(\omega)$ in \eqref{eq: psi} is given by 
\vspace*{-1.5 mm}
\begin{equation}
    \psi_{i,j}^{}(\omega) \; = \; \frac{
    \EX\{ Y_{i}^{}(\omega) Y_{j}^{*}(\omega) \}
    }{
    | \EX\{ Y_{i}^{}(\omega) Y_{j}^{*}(\omega) \} |
    } \; ,
\label{eq: psi}
\vspace*{-2 mm}
\end{equation}
where $Y_{m}(\omega)$ denotes the $m$-th microphone signal in the frequency-domain and $\EX\{\cdot{}\}$ the expectation operator. 
The PHAT weighting in \eqref{eq: psi} has been shown to improve robustness against reverberation and noise \cite{chen2006time, velasco2016proposal, zhang2008does}. 
The TDOA $\hat{\tau}_{i,j}$ between the $i$-th and $j$-th microphone is estimated by maximizing $\xi_{i,j}(\tau)$, i.e., 
\vspace*{-2 mm}
\begin{equation}
    \hat{\tau}_{i,j} \;\; = \;\; \argmax_{\tau} \;\; \xi_{i,j}(\tau) \; .
    \label{eq: TDOA estimation}
\vspace*{-1.5 mm}
\end{equation}

Building upon GCC-PHAT, the SRP-PHAT method \cite{dibiase2000high} is a popular method for 3D source localization. 
The SRP-PHAT functional for the 3D position $\pbf = [p_x, p_y, p_z]^{\textrm{T}}$ is defined as 
\vspace*{-1.2 mm}
\begin{equation}
    \Psi(\pbf) \; = \;\;\;  {\sum_{i,j : i > j}^{} \int_{-\omega_{0}}^{\omega_{0}} \psi_{i,j} (\omega) e^{\ji \omega \tau_{i,j}(\pbf)} }\;\;\; d\omega \; , 
\label{eq: SRP Integral}
\vspace*{-1.5 mm}
\end{equation} 
where $\tau_{i,j}(\pbf)$ denotes the TDOA corresponding to a source at position $\pbf$ and the TDOAs between all microphone pairs are considered. 
The source position is estimated as 
\vspace*{-2 mm}
\begin{equation}
    \hat{\s}_{\textrm{SRP-PHAT}} \;\; = \; \argmax_{ \pbf }  \; \Psi(\pbf) \; ,
    \label{eq: SRP-PHAT}
\vspace*{-1.5 mm}
\end{equation}
which requires the optimization of three continuous variables, i.e., $0 \leq p_{x} \leq P_{x}$, $0 \leq p_{y} \leq P_{y}$ and $0 \leq p_{z} \leq P_{z}$, with $P_{x}$, $P_{y}$ and $P_{z}$ the room dimensions. 

\section{EDM-Based Source Localization}  
\label{sec: EDM-Based Localization}

In this section, we show how to determine the source position, by constructing an EDM of the distances between the microphones and between the source and the microphones (Section \ref{sec: Properties of EDM Matrices}) in a way, which, together with the TDOAs, allows us to build a cost function to determine the (unknown) distance between the source and the reference microphone (Section \ref{sec: EDM-Based Cost Function}). 
Furthermore, we propose a method to select the best TDOA estimate out of multiple candidate estimates (Section \ref{sec: TDOA Candidate Selection}). 

\subsection{Properties of EDM Matrices}
\label{sec: Properties of EDM Matrices}
We define the $3\times(M+1)$-dimensional positions matrix as \linebreak{}$\Pbf = [ \Mbf | \, \s ]$. 
The corresponding $(M+1) \times (M+1)$-dimensional EDM $\overline{\D}$ is defined as 
\vspace*{-2.5 mm}
\begin{equation}
\overline{\D} \; = \; 
\begin{bmatrix}
\begin{array}{c|c}
  \D_{}^{} & \dbf_{}^{} \\
  \hline
  \dbf_{}^{\textrm{T}} & 0_{}^{}
\end{array}
\end{bmatrix} \; .
\label{eq: EDM}
\vspace*{-1.8 mm}
\end{equation}
This matrix contains the inter-microphone EDM $\D_{}^{} = \left[D^{2}_{i,j}\right]$, with $D_{i,j} = ||\m_{i} - \m_{j}||$ the distances between the $i$-th and $j$-th microphones, and the Euclidean distance vector $\dbf = \left[d^{2}_{1}, \dots, d^{2}_{M}\right]^{\textrm{T}}$, with $d_{m} = ||\m_{m} - \s||$ the distance between the source and the $m$-th microphone. 

In \cite{gower1982euclidean, dokmanic2015euclidean}, it was shown that an EDM corresponding to a 3D geometry can be transformed to a Gram matrix, whose rank is at most 3, as 
\vspace*{-1 mm}\begin{equation}
    \G \; = \; -\frac{1}{2}(\I-\1\e^{\textrm{T}})\overline{\D}(\I-\e\1^{\textrm{T}}) \; ,
    \label{eq: EDM to Gram}
    \vspace*{-0.6 mm}
\end{equation}
where $\I$ denotes the identity matrix, $\1$ denotes a vector with ones, and $\e$ denotes a vector with zeros except for the element corresponding to the reference microphone (chosen as the first microphone without loss of generality), equal to one. 
The Gram matrix can be written using the relative microphone and source positions $\Pbf_{\textrm{rel}}^{}$ as $\G = \Pbf_{\textrm{rel}}^{\textrm{T}} \Pbf_{\textrm{rel}}^{}$, where the absolute positions $\Pbf$ are related to the relative positions $\Pbf_{\textrm{rel}}$ via a translation which places the reference microphone at the origin, and the remaining array is arbitrarily rotated and/or reflected (preserving the inter-microphone distances). 
Realizing that the positive semi-definite Gram matrix has at most 3 positive eigenvalues which are not equal to zero, i.e., $\lambda_1 \geq \dotsc \geq \lambda_3 \geq 0$ and $\lambda_{4} = \dots = \lambda_{M+1} = 0$, the relative positions $\Pbf_{\textrm{rel}}$ can be written using the eigenvalue decomposition of $\G$ as 
\vspace*{-1.5 mm}\begin{equation}
    \Pbf_{\text{rel}} = \left[\text{diag}\left(\sqrt{\lambda_1},\dots,\sqrt{\lambda_{3}}\right) \; |\; \0_{{3}\times((M+1)-3)}\right] \U^{\text{T}} \; ,
    \label{eq: Gram EVD}
\vspace*{-1.7 mm}
\end{equation}
where $\0_{{3}\times((M+1)-3)}$ is a $3 \times ((M+1)-3)$ dimensional matrix of zeros and $\U$ denotes the matrix containing the eigenvectors of $\G$. 
The relative positions $\Pbf_{\textrm{rel}}$ can be aligned with the absolute positions $\Pbf$ using orthogonal Procrustes analysis \cite{schoenemann1964solution,dokmanic2015euclidean} by aligning the relative microphone positions $\Mbf_{\textrm{rel}}$ with the known microphone positions $\Mbf$. This simultaneously aligns the relative source position $\s_{\textrm{rel}}$ with the absolute source position $\s$. 

\subsection{EDM-Based Cost Function}
\label{sec: EDM-Based Cost Function}
Defining $\alpha_{s}$ as the distance between the source and the reference microphone (i.e., $\alpha_{s} = d_{1}$), the distance between the source and the $m$-th microphone can be written as
\vspace*{-0.75 mm}
\begin{equation}
    d_{m} \;\; = \;\; \alpha_{s} \; + \; \nu \, \tau_{m,1}(\s) \;\;\, , \;\; \quad m = 1 \;, \; \dots \; , \; M \; ,
    \label{eq: d_MS}
\vspace*{-0.75 mm}
\end{equation}
where $\tau_{m,1}(\s)$ denotes the TDOA between the $m$-th microphone and the reference microphone. 
Assuming for now that the TDOAs are known and considering the distance variable $\alpha$, we can define $d_{m}(\alpha)$ similarly to \eqref{eq: d_MS}, i.e., 
\vspace*{-2.2 mm}\begin{equation}
    d_{m}(\alpha) \;\; = \;\; \alpha \; + \; \nu \, \tau_{m,1}(\s) \;\;\, .
    \label{eq: d_MS(alpha)}
\vspace*{-0.75 mm}
\end{equation}
Using $d_{m}(\alpha)$, we can construct the Euclidean distance vector $\dbf(\alpha) = \left[d^{2}_{1}(\alpha), \dots, d^{2}_{M}(\alpha)\right]^{\textrm{T}}$, the EDM $\overline{\D}(\alpha)$ and its Gram matrix $\G(\alpha)$. 
As mentioned in Section \ref{sec: Properties of EDM Matrices}, the rank of $\G(\alpha)$ is equal to 3 if $\alpha = \alpha_{s}$. 
Motivated by the idea of minimizing the rank of a matrix as in \cite{roy2007effective}, we now define the cost function
\vspace*{-2.3 mm}
\begin{equation}
\boxed{ 
J\left(\alpha\right) = 
\sum_{i=3+1}^{M+1} 
 |\lambda_i\left(\alpha\right)| 
 \;
}
\label{eq: source-ref-distance estimation_}
\vspace*{-1.5 mm}
\end{equation}
which considers all but the three largest eigenvalues $\lambda_{i}(\alpha)$ of $\G(\alpha)$. 
The absolute values of the eigenvalues are used, since it can not be guaranteed that the eigenvalues of $\G(\alpha)$ are positive for all values of $\alpha$ (e.g., in case of a mismatch with the TDOAs). 
If $\alpha = \alpha_{s}$, then all but the three largest eigenvalues of $\G(\alpha_{s})$ are equal to zero, such that $J\left(\alpha_{s}\right) = 0$. 
The optimal value $\alpha_{s}$ can hence be found as 
\vspace*{-0.5 mm}\begin{equation}
\alpha_{s} \; = \; 
\argmin_{\alpha} 
\;\; J\left(\alpha\right) \; .
\label{eq: source-ref-distance estimation}
\vspace*{-1.6 mm}
\end{equation}
\subsection{TDOA Selection}
\label{sec: TDOA Candidate Selection}
\begin{figure}[t!]
\centering
\includegraphics[width=\linewidth]{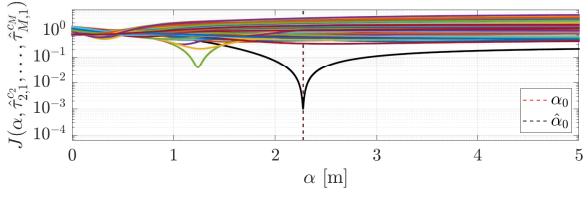}\vspace*{-4 mm}
\caption{Example of the cost function \eqref{eq: TDOA candidates cost} using estimated TDOAs, considering $C=3$ candidate TDOA estimates, with $M=6$ microphones (i.e., $C^{M-1} = 3^{5} = 243$ total combinations), for a distance $\alpha_{s} = 2.28$ m between the source and the reference microphone.}
\label{fig: cost function multiple candidates}
\end{figure}

In practice, the TDOAs, of course, aren't available, so we now rewrite \eqref{eq: d_MS(alpha)} to take into account the estimated TDOAs. 
If the source or a microphone is close to a wall or the corner of a room, super-positions of acoustic reflections may lead to peaks in the time-domain GCC-PHAT function \eqref{eq: IFT of GCC-PHAT}, which are higher than the peak corresponding to the direct path. 
Basing the TDOA estimate on these erroneous peaks can result in large errors in the source localization. 
We propose to consider $C$ candidate TDOA estimates per microphone pair, corresponding to the $C$ highest local peaks in the time-domain GCC-PHAT function. 
The index $c_m \in \{1, \dots, C\}$ denotes the candidate TDOA estimate $\hat{\tau}_{m,1}^{c_m}$ between the $m$-th microphone and the reference microphone. 
This means that the distance variable $d_{m}(\alpha, \hat{\tau}_{m,1}^{c_m})$ now becomes a function of the distance variable $\alpha$ as well as the estimated candidate TDOA estimates, i.e., 
\vspace*{-1.2 mm}\begin{equation}
d_{m}(\alpha, \hat{\tau}_{m,1}^{c_m}) \;\; = \;\; \alpha \; + \; \nu \, \hat{\tau}_{m,1}^{c_m} \; .
    \label{eq: d_est}
\vspace*{-0.6 mm}
\end{equation}
Similarly to Section \ref{sec: EDM-Based Cost Function}, we now construct the Euclidean distance vector $\dbf(\alpha, \hat{\tau}_{2,1}^{c_2}, \dots, \hat{\tau}_{M,1}^{c_M})$, the EDM $\overline{\D}(\alpha, \hat{\tau}_{2,1}^{c_2}, \dots, \hat{\tau}_{M,1}^{c_M})$ and its Gram matrix $\G(\alpha, \hat{\tau}_{2,1}^{c_2}, \dots, \hat{\tau}_{M,1}^{c_M})$, and determine the optimal distance variable $\hat{\alpha}_{s}$ with \eqref{eq: source-ref-distance estimation_} for all $C^{M-1}$ possible combinations of candidate TDOA estimates, taking the value with the minimal cost, i.e., 
\vspace*{-2 mm}
\begin{equation} 
\boxed{
    \hat{\alpha}_{s} \; \; \; \; = \; 
        \;\argmin_{\alpha ,  c_2, \dots, c_M } \;\;
        J(\alpha, \hat{\tau}_{2,1}^{c_2}, \dots, \hat{\tau}_{M,1}^{c_M})
}
\label{eq: TDOA candidates cost}
\vspace*{-0.5 mm}
\end{equation}
This corresponds to the distance between the source and the reference microphone $\alpha$ and the combination of candidate TDOA estimates which best match with each-other in terms of constructing a 3D geometry. 
It should be noted that the minimum of \eqref{eq: TDOA candidates cost} is not guaranteed to be 0 like in \eqref{eq: source-ref-distance estimation_}, due to possible estimation errors in the TDOAs. 
For an exemplary 3D source and microphone constellation with $\alpha_{s} = d_{1} = 2.28$ m, Fig. \ref{fig: cost function multiple candidates} depicts the dependence of the cost function $J(\alpha, \hat{\tau}_{2,1}^{c_2}, \dots, \hat{\tau}_{M,1}^{c_M})$ on the distance variable $\alpha$. 

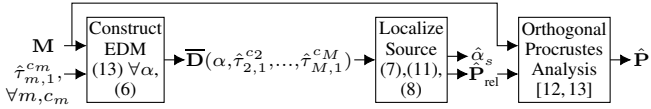
\begin{figure}
\centering
\scriptsize
\begin{tikzpicture}[scale = 0.188]
\node[align=center] at (-3.25,1) {$\Mbf$};
\draw[-{Latex[length=1.5mm, width=1.5mm]}] (-1.8,1) -- (-0.2,1);
\draw[-{Latex[length=1.5mm, width=1.5mm]}] (-1.25,1) -- (-1.25,4) -- (28.5,4) -- (28.5,1) -- (30,1);

\node[align=center] at (-3.6,-1.7) {$\hat{\tau}^{c_m}_{m,1},$\\
$\forall m, c_m$};
\draw[-{Latex[length=1.5mm, width=1.5mm]}] (-1.8,-1) -- (-0.2,-1);
\draw (-0.2,-3) rectangle (5.2,3);
\node[align=center] at (2.5,0) {Construct\\
EDM\\
\eqref{eq: d_est} $\forall \alpha$, \\
\eqref{eq: EDM}}; 
\draw[-{Latex[length=1.5mm, width=1.5mm]}] (5.2,0) -- (6.7,0); 
\node[align=center] at (12.55,-0.12) {$\overline{\D}(\alpha, \hat{\tau}_{2,1}^{c_2}, \dots, \hat{\tau}_{M,1}^{c_M})$};
\draw[-{Latex[length=1.5mm, width=1.5mm]}] (18.5,0) -- (20,0);
\draw (20,-3) rectangle (25,3);
\node[align=center] at (22.5,0) {Localize\\
Source\\
\eqref{eq: EDM to Gram},\eqref{eq: source-ref-distance estimation_},\\
\eqref{eq: Gram EVD}};
\draw[-{Latex[length=1.5mm, width=1.5mm]}] (25,0.25) -- (26.5,0.25);
\node[align=center] at (27.5,0.285) {$\hat{\alpha}_{s}^{}$}; 
\draw[-{Latex[length=1.5mm, width=1.5mm]}] (25,-1) -- (26.5,-1);
\node[align=center] at (27.6,-1) {$\hat{\Pbf}_{\textrm{rel}}^{}$}; 
\draw[-{Latex[length=1.5mm, width=1.5mm]}] (28.5,-1) -- (30,-1);
\draw (30,-3) rectangle (36.4,3);
\node[align=center] at (33.2,0) {Orthogonal\\
Procrustes\\
Analysis\\
\cite{schoenemann1964solution,dokmanic2015euclidean}};
\draw[-{Latex[length=1.5mm, width=1.5mm]}] (36.4,0) -- (37.9,0);
\node[align=center] at (38.65,0) {$\hat{\Pbf}^{}_{}$};
\end{tikzpicture}\vspace*{-4 mm}
\caption{Overview of EDM-Based Source Localization\vspace*{-0.2 cm}} \vspace*{-4 mm} 
\label{fig: Overview}
\end{figure}

To reconstruct the estimated relative positions $\hat{\Pbf}_{\textrm{rel}}$, only the three largest positive eigenvalues (for which the cost function \eqref{eq: TDOA candidates cost} is minimized) and the corresponding eigenvectors are used in \eqref{eq: Gram EVD}. 
The same alignment procedure is applied as described in Section \ref{sec: EDM-Based Localization} to map the estimated relative microphone position $\hat{\s}_{\textrm{rel}}$ to the estimated microphone position $\hat{\s}$. 
An overview of the EDM-based source localization is depicted in Fig. \ref{fig: Overview}. 

\section{Practical Implementation}
\label{sec: Practical Implementation}
In this section, we discuss practical considerations to implement the previously discussed localization algorithms from Sections \ref{sec: SRP-PHAT} and \ref{sec: EDM-Based Localization} in the short-time Fourier transform (STFT) domain. 

\subsection{Implementation of SRP-PHAT} 
\label{sec: Implementation of SRP-PHAT} 
In practice, the maximization of the SRP functional $\Psi(\pbf)$ in \eqref{eq: SRP-PHAT}, which depends on 3 continuous variables, is approximated through an exhaustive search on a discrete grid $\pbf'$. 
First, the integral in \eqref{eq: SRP Integral} is approximated by a sum over STFT frequency bins, i.e.,  i.e., 
\vspace*{-2.3 mm}\begin{equation}
    \Psi[l](\pbf') \;\;\;\;\; = { \sum_{i,j : i > j}^{} \sum_{k=0}^{K-1} 
    \psi_{i,j} [k,l] e^{\ji 2\pi k \tau_{i,j}(\pbf') / K} } \;\; ,
\label{eq: SRP Integral practical}
\vspace*{-1.2 mm}
\end{equation} 
with frequency bin index $k \in \{ 0, \dots, K-1 \}$, with $K$ the Fourier transform length, and frame index $l \in \{ 1, \dots, L \}$. 
Assuming a static source, the source position is then estimated by maximizing the sum of the SRP-PHAT functionals over $L$ frames, i.e., 
\vspace*{-2.2 mm}\begin{equation}
    \hat{\s}'_{\textrm{SRP-PHAT}} \;\; = \; \argmax_{ \pbf' }  \; \sum_{l=1}^{L} \;\; \Psi[l](\pbf') \;\; . 
\label{eq: SRP practical}
\vspace*{-2.1 mm}
\end{equation} 
Since we perform a summation over frames in \eqref{eq: SRP practical}, we use instantaneous estimates of $\psi_{i,j} [k,l]$ in \eqref{eq: SRP Integral practical} (i.e, the expectation operation in \eqref{eq: psi} constitutes an average over a single frame) in order to not perform two temporal averaging operations. 

Exhaustively searching for the 3D source position estimate at a high grid resolution can be computationally demanding. 
Loosely based on coarse-to-fine region contraction in \cite{do2007fast}, we first evaluate \eqref{eq: SRP practical} on a coarse 3D grid and then in the vicinity of a few points where SRP-PHAT yields the highest values, we evaluate the SRP-PHAT functional on a fine grid in those regions. 

\subsection{Implementation of EDM-Based Localization} 
\label{sec: Implementation of EDM-Based Localization} 
In the STFT-domain, GCC-PHAT $\psi_{i,j}[k,l]$ is estimated in each frequency bin $k$ and time frame $l$, and the the continuous Fourier transform in time-domain GCC-PHAT is approximated with an inverse discrete Fourier transform for discrete time-lags $n$, i.e., 
\vspace*{-2.2 mm}\begin{equation}
    \xi_{m,1}[n,l] \;\;\;\;\; = \sum_{k=0}^{K-1} \psi_{m,1}^{}[k,l] e^{\ji 2\pi k n / (f_s K)} \;\; .
    \label{eq: IDFT of GCC-PHAT}
\vspace*{-1.8 mm}
\end{equation} 
To achieve a more precise TDOA estimate, the time-domain GCC-PHAT function $\xi_{m,1}[n,l]$ can be interpolated between the discrete time-lags $n$ with a factor $R \geq 1$. 
The lower and upper limits of the possible time-lags are dependent on the distances between the microphone pair, i.e., $ | n_m | < D_{m,1} f_{s} R / c $. 
Using the interpolated time-domain GCC-PHAT, the sample-delay between the $m$-th and the reference microphone is estimated as 
\vspace*{-2.8 mm}\begin{equation}
    \hat{n}_m \;\; = \;\; \argmax_{n_m} \;\; \sum_{l=1}^{L}\xi_{m,1}[n_m,l] \;\; ,
    \label{eq: discrete TDOA estimation}
\vspace*{-2 mm}
\end{equation} 
corresponding to the estimated TDOA $\hat{\tau}_{m,1}^{} = \hat{n}_m / (f_s R)$. 
Similarly to the previous section, since we perform a summation over frames in \eqref{eq: discrete TDOA estimation}, we use instantaneous estimates of $\psi_{i,j} [k,l]$ in \eqref{eq: IDFT of GCC-PHAT}. 

\section{Experimental Evaluation}
\label{sec:majhead}
In this section, we experimentally compare the source localization performance of the proposed EDM-based method (for up to three candidate TDOA estimates $C$ per microphone pair) with the SRP-PHAT method for four different distances between the source position and the centroid of the microphone positions. 

\begin{figure*}[t!]
    \centering
    \hspace*{1.45 cm}\includegraphics[width=0.85\linewidth]{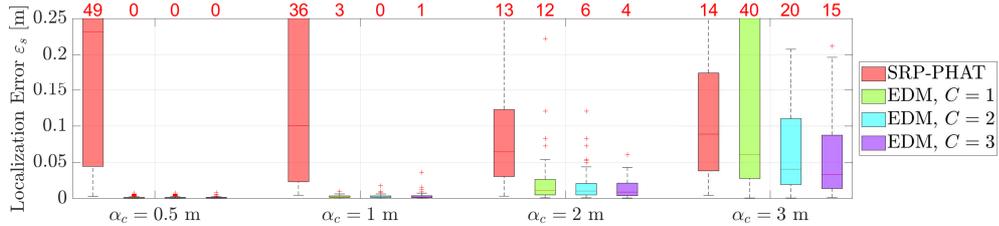}
	\caption{Box plots of the localization errors $\varepsilon_{s}$ (over 100 scenarios) for the SRP-PHAT method and the EDM-based method (with different numbers of candidate TDOA estimates $C$ per microphone pair), for different distances $\alpha_{c}$ between the source and the centroid of the distributed microphones. 
	The number of results outside of the plotted range are denoted by red numbers at the top.} \vspace*{-1.5 mm} 
	\label{fig: Plot1}
\end{figure*}

\subsection{Scenario and Algorithm Parameters}
\label{sec: Scenario}
For the simulations, we considered a rectangular room with dimensions $6\times{}6\times{}2.4$ m and simulated room impulse responses using the image method \cite{HabetsRIR, allen1979image}, assuming equal reflection coefficients for all walls. 
The $M=6$ spatially distributed microphones were randomly positioned within a cube with cube length 2 m (with a minimum distance of 2 cm between the microphones) and the source was located at one of four fixed distances $\alpha_{c} \in \{0.5, 1, 2, 3\}$ m from the centroid of the microphone positions (in a random direction). 
For each source distance $\alpha_{c}$, we considered 100 acoustic scenarios, using a 5 s speech signal randomly selected from \cite{M-AILABS} (with equal probability for a male or female speaker) as the source signal. 
The reflection coefficients were set for each scenario such that the room impulse responses had an average direct-to-reverberant ratio of approximately $0$ dB over the microphones. 
This was achieved by setting $T_{60} = 0.60 \pm 0.14$ s for $\alpha_{c} = 0.5$ m, $T_{60} = 0.47 \pm 0.10$ s for $\alpha_{c} = 1$ m, $T_{60} = 0.29 \pm 0.04$ s for $\alpha_{c} = 2$ m and $T_{60} = 0.25 \pm 0.03$ s for $\alpha_{c} = 3$ m. 
Spherically isotropic multi-talker babble noise was generated using \cite{habets2008generating} and added to the reverberant speech component in the microphones at 5 dB signal-to-noise ratio. 
The sampling frequency was equal to 16 kHz. 

The algorithms were implemented using an STFT framework, with a frame length of $512$ samples (corresponding to 32 ms), 50\% overlap between frames, a discrete Fourier transform-length of $1024$ samples and using a square-root-Hann analysis window. 

For SRP-PHAT, the functional in \eqref{eq: SRP Integral practical} was evaluated first on a coarse grid with 10 cm resolution in x-, y-, and z-direction, and then reevaluated for the three grid points with the highest SRP-PHAT value on a fine grid with 1 cm resolution in each dimension. 
For the proposed EDM-based source localization method, the time-domain GCC-PHAT function in \eqref{eq: IDFT of GCC-PHAT} was interpolated by a factor $R=720$. 
To emphasize strong peaks, we weighted the time-domain GCC-PHAT function as $\widetilde{\xi}_{m,1}[n_{m},l] = \exp (15 \xi_{m,1}[n_{m},l])$ prior to the TDOA estimation in \eqref{eq: discrete TDOA estimation}. 
The candidate TDOAs were selected using a peak finding algorithm \cite{findpeaks}. 
In \eqref{eq: TDOA candidates cost}, the exhaustive search for the optimal distance variable $\alpha_{s}$ was performed with a resolution of 1 mm, up to a maximal distance determined by the distance between opposite corners of the room (i.e., $\sqrt{6^2 + 6^2 + 2.4^2}\textrm{ m} \; \approx \; 8.82\textrm{ m}$). 
For $C=2$ candidate TDOA estimates per microphone pair, the number of combinations of TDOA estimates was $C^{M-1} = 2^{5} = 32$, while for $C=3$ the number of combinations was $C^{M-1} = 3^{5} = 243$. 

\subsection{Performance Comparison}
\label{sec: Influence of candidates and weighting}
\begin{table}[t]
    \vspace*{-2 mm}\centering
    \caption{Median localization errors (over 100 scenarios) for the SRP-PHAT method and the EDM-based method, corresponding to the box plots in Fig. \ref{fig: Plot1}}\vspace*{1 mm}
    \scriptsize
    \begin{tabular}{| c || c | c | c | c |}
        \hline
        &  \multicolumn{4}{c|}{Median $\varepsilon_{s}$ [m]} \\
        $\alpha_{c} [m]$ & SRP-PHAT$^{}$ & EDM, $C=1^{}$ & EDM, $C=2^{}$ & EDM, $C=3^{}$ \\\hline\hline
        $0.5$ & 0.231 & \textbf{0.001} & \textbf{0.001} & \textbf{0.001} \\\hline
        $1$ & 0.101 & \textbf{0.002} & \textbf{0.002} & \textbf{0.002} \\\hline
        $2$ & 0.065 & 0.011 & 0.010 & \textbf{0.009} \\\hline
        $3$ & 0.089 & 0.061 & 0.041 & \textbf{0.033}\\\hline
    \end{tabular}
    \label{tab:my_label}\vspace*{-1.5 mm}
\end{table}
To analyze and compare the performance of the considered 3D source localization methods, we used the localization error 
\vspace*{-1.2 mm}
\begin{equation}
    \varepsilon_{s} \;\; = \;\; ||\s \; - \; \hat{\s}|| \; .
    \vspace*{-1.2 mm}
\end{equation}
For different source distances $\alpha_{c}$, Fig. \ref{fig: Plot1} depicts the box plots of the localization error (over 100 scenarios) for the SRP-PHAT method and for the proposed EDM-based method, for different numbers of candidate TDOA estimates. 
Tab. \ref{tab:my_label} presents the corresponding median localization errors. 

Considering source positions outside of the array of distributed microphones (i.e., $\alpha_{c} \geq 2$ m), it is clear from Fig. \ref{fig: Plot1} that by increasing the number of candidate TDOA estimates, both the median localization errors as well as the number of errors larger than 25 cm are reduced. 
This suggests that the proposed procedure, considering multiple candidate TDOA estimates, is able to identify the TDOA corresponding to the direct path. 
Using $C=2$ or $C=3$ suffices, to halve the number of localization errors larger than 25 cm, compared to using $C=1$, and the median localization error can be substantially reduced, especially for large source distances $\alpha_{c}$. 
For source positions within the array of distributed microphones (i.e., $\alpha_{c} \leq 1$ m),\linebreak{} considering more than $C=1$ candidate TDOA estimates is not necessary, since the median localization error (in Tab. \ref{tab:my_label}) is constantly at 1 mm or 2 mm, independently of $C$. 

In Fig. \ref{fig: Plot1} it can clearly be observed that when the distance between the source and the centroid of the microphones $\alpha_{c}$ is smaller than or equal to the cube length of the array of distributed microphones (i.e., $\alpha_{c} \leq 2$ m), the proposed EDM-based source localization method results in significantly lower localization errors than the SRP-PHAT method, regardless of the number of candidate TDOA estimates. 
For example, for $\alpha_{c} = 2$ m, the median localization error for the EDM-based method is 1 cm $\pm$1 mm (depending on $C$), whereas for the SRP-PHAT method the median localization error is 6.5 cm. 

For $\alpha_{c} = 3$ m, the EDM-based method with $C=3$ candidate TDOA estimates and the SRP-PHAT method have overlapping distributions of localization errors and a comparable number of errors larger than 25 cm, but the EDM-based method has a lower median error, i.e., 3.3 cm, compared to 8.9 cm for the SRP-PHAT method. 

\section{Conclusions and Outlook} 
\label{sec:page}
We have proposed a new 3D source localization method, which, through properties of EDMs, and by the specific construction of the EDM containing the distances between microphones and between the source and the microphones, only requires the optimization of a single variable, namely the distance between the source and the reference microphone. 
As this method relies on estimated TDOAs, we proposed a method to select the best TDOA estimate out of multiple estimates in the presence of reverberation. 
Experimental results for different source and microphone constellations showed that the proposed EDM-based source localization method consistently localizes sources with a lower localization error than the commonly used SRP-PHAT method for all tested source distances. 
The proposed method for estimating the best candidate TDOA estimates also results in a reduction in the number of large localization errors. 

The EDM-based source localization method is currently being adapted for far field sources, multiple sources, and moving sources.

\newpage

\bibliographystyle{IEEEbib}
\balance
\bibliography{ms}

\begin{thebibliography}{10}

\bibitem{dibiase2000high}
J~H DiBiase,
\newblock {\em A high-accuracy, low-latency technique for talker localization
  in reverberant environments using microphone arrays},
\newblock Ph.D. thesis, Brown University, Providence, RI, USA, 2000.

\bibitem{madhu2008acoustic}
N~Madhu, R~Martin, U~Heute, and C~Antweiler,
\newblock ``Acoustic source localization with microphone arrays,''
\newblock {\em Advances in Digital Speech Transmission}, pp. 135--170, 2008.

\bibitem{huang2008time}
Y~A Huang, J~Benesty, and J~Chen,
\newblock ``Time delay estimation and source localization,''
\newblock in {\em Springer Handbook of Speech Processing}, pp. 1043--1063.
  Springer, 2008.

\bibitem{pertila2018multichannel}
P~Pertil{\"a}, A~Brutti, P~Svaizer, and M~Omologo,
\newblock ``Multichannel source activity detection, localization, and
  tracking,''
\newblock {\em Audio source separation and speech enhancement}, pp. 47--64,
  2018.

\bibitem{knapp1976generalized}
C~Knapp and G~Carter,
\newblock ``The generalized correlation method for estimation of time delay,''
\newblock {\em IEEE Trans. on Audio, Speech, Language Processing}, vol. 24, no.
  4, pp. 320--327, 1976.

\bibitem{do2007fast}
H~Do and H~F Silverman,
\newblock ``A fast microphone array \uppercase{SRP-PHAT} source location
  implementation using coarse-to-fine region contraction (\uppercase{CFRC}),''
\newblock in {\em Proc. IEEE Workshop on Applications of Signal Processing to
  Audio and Acoustics (WASPAA)}, New Paltz, NY, USA, 2007, pp. 295--298.

\bibitem{cobos2010modified}
M~Cobos, A~Marti, and J~J Lopez,
\newblock ``A modified \uppercase{SRP-PHAT} functional for robust real-time
  sound source localization with scalable spatial sampling,''
\newblock {\em IEEE Signal Processing Letters}, vol. 18, no. 1, pp. 71--74,
  2010.

\bibitem{nunes2014steered}
L~O Nunes, W~A Martins, M~V~S Lima, L~W~P Biscainho, M~V~M Costa, F~M
  Gon{\c{c}}alves, A~Said, and B~Lee,
\newblock ``A steered-response power algorithm employing hierarchical search
  for acoustic source localization using microphone arrays,''
\newblock {\em IEEE Trans. on Signal Processing}, vol. 62, no. 19, pp.
  5171--5183, 2014.

\bibitem{garcia2021analytical}
G~Garc{\'\i}a-Barrios, J~M Guti{\'e}rrez-Arriola, N~S{\'a}enz-Lech{\'o}n, V~J
  Osma-Ruiz, and R~Fraile,
\newblock ``Analytical model for the relation\linebreak{} between signal
  bandwidth and spatial resolution in steered-response power phase transform
  (\uppercase{SRP-PHAT}) maps,''
\newblock {\em IEEE Access}, vol. 9, pp. 121549--121560, 2021.

\bibitem{dietzen2021low}
T~Dietzen, E~De~Sena, and T~Van~Waterschoot,
\newblock ``Low-complexity steered response power mapping based on
  \uppercase{N}yquist-\uppercase{S}hannon sampling,''
\newblock in {\em Proc. IEEE Workshop on Applications of Signal Processing to
  Audio and Acoustics (WASPAA)}, New Paltz, NY, USA, 2021, pp. 206--210.

\bibitem{torgerson1952multidimensional}
W~S Torgerson,
\newblock ``Multidimensional scaling: I. theory and method,''
\newblock {\em Psychometrika}, vol. 17, no. 4, pp. 401--419, 1952.

\bibitem{dokmanic2015euclidean}
I~Dokmanic, R~Parhizkar, J~Ranieri, and M~Vetterli,
\newblock ``Euclidean distance matrices: essential theory, algorithms, and
  applications,''
\newblock {\em IEEE Signal Processing Magazine}, vol. 32, no. 6, pp. 12--30,
  2015.

\bibitem{schoenemann1964solution}
P~Schoenemann,
\newblock {\em A solution of the orthogonal Procrustes problem with
  applications to orthogonal and oblique rotation},
\newblock Ph.D. thesis, University of Illinois, Urbana-Champaign, 1964.

\bibitem{velasco2016proposal}
J~Velasco, C~J Martin-Arguedas, J~Macias-Guarasa, D~Pizarro, and M~Mazo,
\newblock ``Proposal and validation of an analytical generative model of
  \uppercase{SRP-PHAT} power maps in reverberant scenarios,''
\newblock {\em Signal Processing}, vol. 119, pp. 209--228, 2016.

\bibitem{zhang2008does}
C~Zhang, D~Flor{\^e}ncio, and Z~Zhang,
\newblock ``Why does \uppercase{PHAT} work well in low noise, reverberative
  environments?,''
\newblock in {\em Proc. IEEE International Conference of Acoustics, Speech and
  Signal Processing (ICASSP)}, Las Vegas, NV, USA, 2008, pp. 2565--2568.

\bibitem{chen2006time}
J~Chen, J~Benesty, and Y~Huang,
\newblock ``Time delay estimation in room acoustic environments: An overview,''
\newblock {\em EURASIP Journal on Applied Signal Processing}, pp. 1--19, 2006.

\bibitem{gower1982euclidean}
J~C Gower,
\newblock ``Euclidean distance geometry,''
\newblock {\em Math. Sci}, vol. 7, no. 1, pp. 1--14, 1982.

\bibitem{roy2007effective}
O~Roy and M~Vetterli,
\newblock ``The effective rank: A measure of effective dimensionality,''
\newblock in {\em Proc. European Signal Processing Conference (EUSIPCO)},
  Poznan, Poland, 2007, pp. 606--610.

\bibitem{HabetsRIR}
E~A~P Habets,
\newblock {\em RIR-Generator},
\newblock Available at \url{https://github.com/ehabets/RIR-Generator}.

\bibitem{allen1979image}
J~B Allen and D~A Berkley,
\newblock ``Image method for efficiently simulating small-room acoustics,''
\newblock {\em Journal of the Acoustical Society of America}, vol. 65, no. 4,
  pp. 943--950, 1979.

\bibitem{M-AILABS}
I~Solak,
\newblock {\em M-AILABS Speech Dataset},
\newblock Available at
  \linebreak{}\url{https://www.caito.de/2019/01/03/the-m-ailabs-speech-dataset/}.

\bibitem{habets2008generating}
E~A~P Habets, I~Cohen, and S~Gannot,
\newblock ``Generating nonstationary multisensor signals under a spatial
  coherence constraint,''
\newblock {\em Journal of the Acoustical Society of America}, vol. 124, no. 5,
  pp. 2911--2917, 2008.

\bibitem{findpeaks}
{\em Matlab \texttt{findpeaks} function},
\newblock Documentation available at
  \url{https://www.mathworks.com/help/pdf_doc/signal/signal_ref.pdf}, p. 414,
  2022.

\end{thebibliography}

\end{document}